\documentclass[twocolumn,showpacs,preprintnumbers,aps,prl,amsmath,amssymb]{revtex4}

\usepackage[english]{babel}
\usepackage[latin2]{inputenc}
\usepackage[T1]{fontenc}

\usepackage[usenames]{color}

\usepackage{ae,aecompl}
\usepackage{enumerate}
\usepackage{epsfig}
\usepackage{float}
\usepackage{graphics}
\usepackage{graphicx}
\usepackage{amsmath}
\usepackage{amssymb}
\usepackage{pifont}
\usepackage[sort&compress]{natbib}
\usepackage{wasysym}
\usepackage{color}

\def\ev #1{\left\langle #1 \right\rangle}

\def\Br #1{\left[ #1 \right]}

\begin{document}
\noindent {\bf  Comment on ``Tests of scaling and universality..." by Plerou and Stanley, Phys. Rev. E {\bf 76},046109}
\\

In a pioneering study Plerou \emph{et al.} [1] extended the statistical analysis of financial data on the distribution of the volume $q$, the number of shares exchanged in a single transaction. They concluded that this decays asymptotically as a power law, $P(q>x) \propto x^{-\zeta_q}$ with $\zeta_q \approx 1.5$, a value in the Levy stable regime. A later study in Ref. [2], however, found significantly higher values for data aggregated over 15 minutes, typically $\zeta_Q \approx 2.3$, outside the Levy regime.

In a recent Article [3], Plerou and Stanley extended their analysis and found $\zeta_q \approx 1.49-1.65$ for three different stock markets. Ref. [3] argued that the discrepancy between the previous works [1,2] in determining the tail exponents
% $\zeta_q$ 
could be traced back to the use of the Hill's estimator (HE) [4], which contains a cutoff parameter that is usually determined in a qualitative fashion. In contrast, Ref. [3] uses the Meerschaert-Scheffler estimator (MSE) [5], which is cutoff and shift independent and defined as
\begin{equation}
	\zeta_q^\mathrm{MS} = \frac{2(\gamma+\ln N)}{\gamma + \ln_+\sum_{i=1}^N(q_i-\ev{q})^2}.
\end{equation}
$q_i$ are independent observations of $q$, and $\ev{q}$ is its mean, $\gamma = 0.5772$ is
Euler's constant and $\ln_+(x) = \max(0, \ln x)$.

The use of the MSE raises several problems in this specific case. First of all, the MSE is an estimator of the tail-index only if the index is less than 2, otherwise (i.e., if the variance is finite) it converges to 2.0. In addition, the rate of convergence is quite slow compared to HE. These two factors result in the MSE being an questionable tool to distinguish between the cases $\zeta > 2$ and $\zeta < 2$. Furthermore, this estimator is not invariant to changes of the scale: for any fixed $A>1$, $\zeta^{\mathrm{MS}}_{q/A} > \zeta^{\mathrm{MS}}_q$. In [3] the data are rescaled with the mean absolute deviation. 

In order to demonstrate that the MSE based procedure yields misleading results, we used it on computer-generated Pareto distributed data with different tail exponents. Sample sizes were chosen to be comparable with that of financial data considering liquid stocks on NYSE in the period studied in Ref. [3] (1994-1995), where $\zeta_q = 1.65 \pm 0.01$ was concluded. Our results are shown in Fig. 1.

The HE has no limitation for the value of the tail exponent to be determined, however, as mentioned in [3], the estimation depends on the cutoff, at least in the original version [4]. In a recent improvement [6] this problem was resolved by using a self-consistent optimization scheme. We have found that the problem of shift-dependence [2] could also be settled in a similar fashion. Using the im-
\begin{figure}[H]
\centerline{\includegraphics[width=230pt]{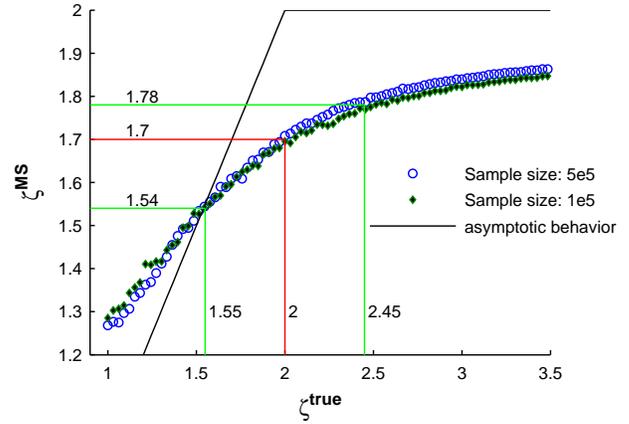}}
\caption{(color online) MSE estimates of the tail exponent of generated Pareto distributed data. Each point corresponds to the average of 100 runs. The sample sizes are indicated in the legend. The solid black line shows the asymptotic ($N \rightarrow \infty$) behavior of the MSE. The empirical value of $\zeta_q=2.02$ with the errors as calculated from the self consistent Hill estimator and the corresponding MSE values are indicated (green and red lines).} 
\label{fig:rwm_internal_scaling}
\end{figure}

\noindent proved Hill schemes we determined the tail exponents for the volume distribution of transactions for the 1000 most liquid stocks on the NYSE for the same period as studied in [3]. The obtained results are significantly higher than those of [3]. We find an average exponent $\zeta_q=2.02\pm 0.45$. The corresponding values are shown in Fig. 1 and it is indicated how such measured values result in smaller MSE estimates. For the aggregated data with a time window of 15 min we find $\zeta_Q \approx 2.3$, in agreement with [2].

Support by OTKA T049238 is acknowledged.

{\raggedleft {\'Eva R\'acz, Zolt\'an Eisler and J\'anos Kert\'esz}\\
{Institute of Physics}\\
{Budapest University of Technology and Economics}\\
{Budapest, Budafoki \'ut 8, H-1111 Hungary}\\
}
\bigskip
\noindent
References:\\
\noindent 
$\Br 1$ P. Gopikrishnan, V. Plerou, X. Gabaix, H. E. Stanley, Phys. Rev. E {\bf 62}, 4493 (2000)\\
$\Br 2$ Z. Eisler, J. Kert\'esz, Eur. Phys. J. B {\bf 51}, 145 (2006)\\
$\Br 3$ V. Plerou, H. E. Stanley, Phys Rev. E {\bf 76}, 046109 (2007)\\
$\Br 4$ B. M. Hill, Ann. Stat. {\bf 3}, 1163 (1975)\\
$\Br 5$ M. M. Meerschaert, H. Scheffler, J. Stat. Plan. Infer. {\bf 71}, 19 (1998); K. Bianchi, M. M. Meerschaert, working paper (2004)\\ http://wolfweb.unr.edu/homepage/mcubed/KBest.pdf\\
$\Br 6$ A. Clauset, C. R. Shalizi, M. E. J. Newman, arXiv:0706.1062v1

\end {document}